\documentclass{IEEEtran}
\usepackage{color}
\usepackage{graphics} % for pdf, bitmapped graphics files
\usepackage{epsfig} % for postscript graphics files
\usepackage{mathptmx} % assumes new font selection scheme installed
\usepackage{times} % assumes new font selection scheme installed
\usepackage{amsmath} % assumes amsmath package installed
\usepackage{amssymb}  % assumes amsmath package installed
\usepackage{graphicx}
\usepackage{multirow, makecell}
\usepackage{array}
\usepackage{booktabs} % For Table
\usepackage{comment}
\usepackage[linesnumbered,ruled]{algorithm2e} % For pseudocode and algorithms
\newcolumntype{M}[1]{>{\centering\arraybackslash}m{#1}}

\title{\vspace{1.0cm}
Vehicle Energy Dataset (VED), A Large-scale Dataset for Vehicle Energy Consumption Research}

\author{G. S. Oh, David J. Leblanc, and Huei Peng
\thanks{
This research was supported by U.S. Department of Energy under the award DE-EE0007212.
}
\thanks{G. S. Oh is with the Department of Mechanical Engineering, University of Michigan, Ann Arbor, MI 48109 USA, Email: gsoh@umich.edu
}
\thanks{David J. LeBlanc is with the University of Michigan Transportation Research Institute, Ann Arbor, MI 48109-2150 USA, Email: leblanc@umich.edu}
\thanks{Huei Peng is the Director of M-city and also a Professor in the Department of Mechanical Engineering, University of Michigan, Ann Arbor, MI 48109 USA, Email: hpeng@umich.edu}
}

\begin{document}

\maketitle

\begin{abstract}
We present Vehicle Energy Dataset (VED), a novel large-scale dataset of fuel and energy data collected from 383 personal cars in Ann Arbor, Michigan, USA. This open dataset captures GPS trajectories of vehicles along with their time-series data of fuel, energy, speed, and auxiliary power usage. A diverse fleet consisting of 264 gasoline vehicles, 92 HEVs, and 27 PHEV/EVs drove in real-world from Nov, 2017 to Nov, 2018, where the data were collected through onboard OBD-II loggers. Driving scenarios range from highways to traffic-dense downtown area in various driving conditions and seasons. In total, VED accumulates approximately 374,000 miles. We discuss participant privacy protection and develop a method to de-identify personally identifiable information while preserving the quality of the data. After the de-identification, we conducted case studies on the dataset to investigate the impacts of factors known to affect fuel economy and identify energy-saving opportunities that hybrid-electric vehicles and eco-driving techniques can provide. The case studies are supplemented with a number of examples to demonstrate how VED can be utilized for vehicle energy and behavior studies. Potential research opportunities include data-driven vehicle energy consumption modeling, driver behavior modeling, machine and deep learning, calibration of traffic simulators, optimal route choice modeling, prediction of human driver behaviors, and decision making of self-driving cars. We believe that VED can be an instrumental asset to the development of future automotive technologies. The dataset can be accessed at https://github.com/gsoh/VED.

\end{abstract}


\section{Introduction}

\IEEEPARstart{C}{onnected} and automated vehicle (CAV) technologies offer new opportunities to reduce energy consumption through different avenues \cite{ref:Review_CAV_Vahidi}, \cite{ref:Impact_CAV1}, \cite{ref:Impact_CAV2} \cite{ref:Impact_CAV3}. Developments of such technologies can benefit from studies on the fuel and energy consumption behavior of human drivers. While there exists a multitude of publicly available datasets for machine vision, radar, and LiDAR \cite{ref:Dataset_KITTI}, \cite{ref:Dataset_BDD100k}, \cite{ref:Dataset_MapVistas}, a large-scale dataset which contains time-series fuel and energy data is not publicly available to the best of our knowledge. 

While comprehensive reports have been published by U.S. Department of Energy (DOE) and U.S. Environmental Protection Agency (EPA) on vehicle fuel economy, the reported results are based largely on standard lab-tests on predefined driving cycles \cite{ref:FEG2017_EPA}, \cite{ref:FCvsSpd_OakRidge}, \cite{ref:AFDC_FuelInfo}. It was widely observed that significant difference exists between the fuel economy obtained in lab tests and in real-world driving \cite{ref:FC_LabtestVSRealworld}, \cite{ref:RealWorldBus_Beijing}, \cite{ref:RealWorldPassenger_Europe}, \cite{ref:ICEvsHEV_FC_Zahabi2014}, \cite{ref:ICEvsEV_FC_China}. In real-world, vehicle energy consumption is affected by various other factors including other road users, traffic signals, road conditions, road grade, heating and ventilation.

A number of real-world traffic data are available to the public \cite{ref:FHWA_HighwayStats2016}. The Federal Highway Administration (FHWA) provides segment level vehicle travel data from Highway Performance Monitoring System (HPMS) \cite{ref:FHWA_HPMS_Manual} and hourly fixed point vehicle count data from Travel Monitoring Analysis System (TMAS) \cite{ref:FHWA_TMS_Guide}. However, they focus on traffic monitoring, providing aggregate traffic information such as traffic volume and vehicle counts, which are typically obtained through loop detectors and visual sensors. The data do not include trajectories of individual vehicles, or time-series energy data. 

To the author's knowledge, a large-scale comprehensive dataset which contains individual vehicles trajectories and time-series energy consumption data does not yet exist in the public domain. In order to bridge this gap, we publish our dataset, the Vehicle Energy Dataset (VED), to the broad public, providing fuel and energy consumption time-series data of various personal cars operated in diverse real-world driving conditions over the time span of a year (Nov, 2017 to Nov, 2018), under the support of DOE.
		
The remainder of this paper is organized as follows: 
Section II introduces the VED by describing the data collection process and providing analysis of the recorded signals; Section III describes how privacy of the vehicles is protected; Section IV elaborates on examples to demonstrate how VED can be utilized to understand fuel consumption of gasoline and hybrid vehicles; Section V describes two case-studies on hybrid and electric vehicles to demonstrate how such studies can benefit from the dataset; Section VI discusses eco-driving techniques and how the dataset can improve eco-driving techniques; finally, Section VII offers concluding remarks.

\section{Vehicle Energy Dataset (VED)}

In this section, we elaborate on the data collection process, the data contents, statistics on the fleet, and the fuel consumption estimation method.

\begin{figure}[t] 
    \centering
    \includegraphics[width=0.95\linewidth]{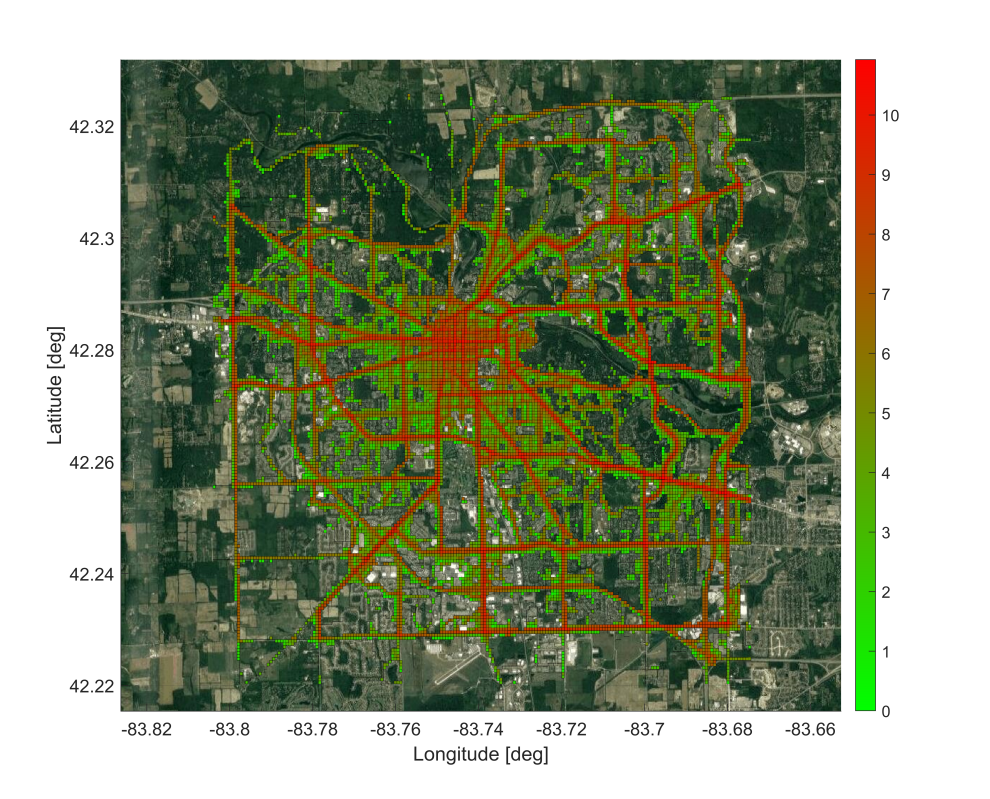}
    \caption{Heatmap of the collected data. The intensity of the heatmap shows natural log of the data density. A grid with the intensity 10 contains $e^{10}$ times more data compared to a grid with the intensity 0.}
    \label{Fig:GPS_Satellite_ColorMap}
\end{figure}

\subsection{Data Collection}

A collaborative effort between the University of Michigan (UM), the Argonne National Lab, and the Idaho National Lab was carried out to study energy consumption behavior of human drivers and energy-saving potentials of eco-driving technologies. As a part of the effort, a large participant pool of personal vehicles owners living and/or working in Ann Arbor, Michigan was recruited and on-board diagnostics II (OBD-II) data loggers were installed into their vehicles with their consents. The installed devices collected driving records over a year, sending real-time data including their GPS trajectories, and OBD-II signals to a database over cellular connections.

OBD-II provides information on the status of the various vehicle subsystems, especially ones related to vehicle emission, via a standardized communication protocol \cite{ref:SAE_OBD2}. The OBD-II port provides an entry point to the vehicle Controller Area Network (CAN), enabling CAN signals to be monitored and recorded. Since all passenger vehicles sold in U.S. were equipped with OBD-II starting in 1996, we were able to collect a consistent set of data across various vehicles.

\begin{table}[h]
\caption{Number of Vehicles and Mileage of The Fleet}
\label{Table:VehicleType}
\begin{tabular}{
>{\centering\arraybackslash}m{1cm}  
>{\centering\arraybackslash}m{2cm} 
>{\centering\arraybackslash}m{2cm} 
>{\centering\arraybackslash}m{2cm} }
\toprule
\multicolumn{2}{c}{\textbf{Vehicle Type}} & \textbf{\# of Vehicles} & \textbf{Mileage}  \\
\toprule
\multirowcell{2}{\\ ICE \\} & Fuel & 249 & 199,331 \\
\cmidrule{2-4}
& No Fuel & 15 & 14,136 \\
\midrule
\multirowcell{2}{\\ HEV \\} & Fuel & 90 & 109,840 \\
\cmidrule{2-4}
& No Fuel & 2 & 466 \\
\midrule
\multicolumn{2}{c}{PHEV} & 24 & 45,470 \\
\midrule
\multicolumn{2}{c}{EV} & 3 & 4,721 \\
\midrule
\multicolumn{2}{c}{\textbf{Total}} & \textbf{383} & \textbf{373,964} \\
\bottomrule
\end{tabular}
\end{table}

In 2016, the market share for hybrid electric cars (HEVs) was about 2\% and the combined market share of plug-in hybrid electric vehicles (PHEVs) and electric vehicles (EVs) was about 1\%. Since mass-produced PHEVs and EVs were available in the United States starting in 2010, their cumulative market share is much lower than their annual share \cite{ref:Veh_MarketShare}. Taking into consideration the relative rarity of these vehicles, efforts were made to recruit individuals with HEVs, PHEVs, and EVs to ensure sufficient data collection for vehicles other than those with only internal combustion engines (ICEs). The result of the recruitment as well as the total mileage of the vehicles in the dataset ("the fleet") is summarized in Table.~\ref{Table:VehicleType}. 

The dataset contains data collected from 383 vehicles from Nov 1st, 2017 to Nov 9th, 2018. The diverse fleet includes passenger cars (coupes, hatchbacks, sedans, convertibles, crossovers, and luxury cars) and light trucks (pickup trucks, SUVs, minivans, and wagons). VED includes a total of 373,964 miles of naturalistic driving data from the fleet; represented as csv files, the data are 2.78 GB in size (uncompressed) and 0.17 GB (compressed). 

\begin{figure}[t] 
    \centering
    \includegraphics[width=0.90\linewidth]{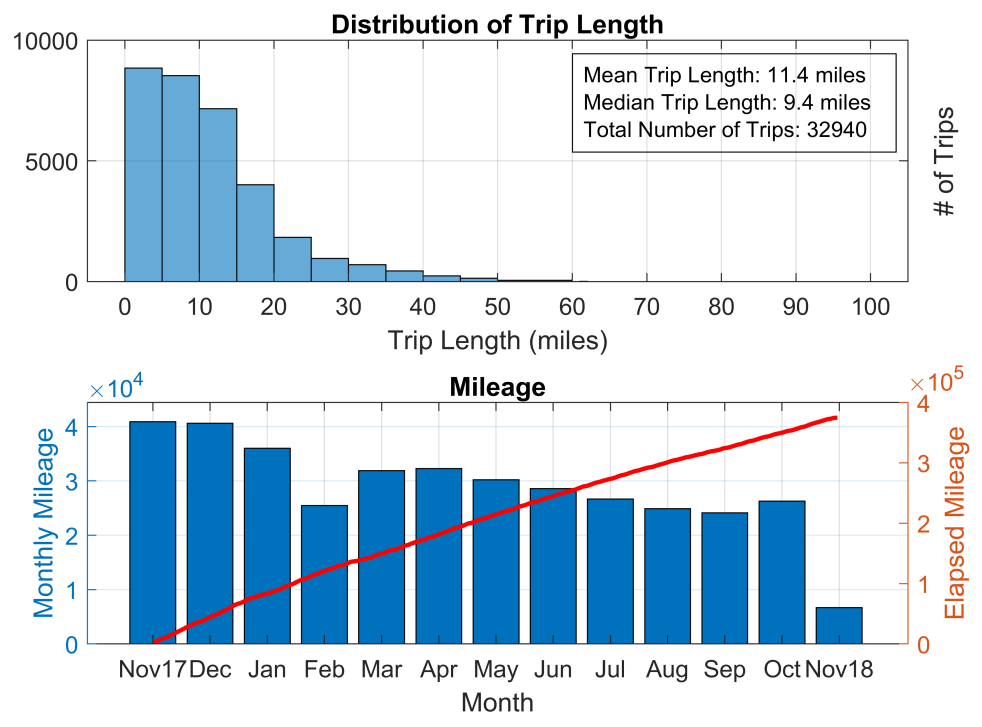}
    \caption{Top plot describes the distribution of trip length, and the bottom plot presents monthly and cumulative mileage the fleet traveled.}
    \label{Fig:Trip_Stats}
\end{figure}

Table.~\ref{Table:VehicleType} provides details on the number of vehicles and total mileages each group accumulated. Among total 373,964 miles traveled by the fleet, 86.6\% and 13.4\% of them are from ICE \& HEVs and PHEVs \& BEVs. Fuel data including fuel-rate, mass air flow, and absolute load are available for most of the vehicles, however, a small portion of the fleet do not provide the fuel data - these are labeled as "No Fuel" in the relevant tables. Fig.~\ref{Fig:GPS_Satellite_ColorMap} illustrates a geographic heatmap for the traces of the collected data, and Fig.~\ref{Fig:Trip_Stats} provides a break-down of the mileage and a distribution of trip length.

\subsection{Data Description}

\begin{table*}[h]
\caption{Contents of Time-stamped Dynamic Data}
\label{Table:Data_Contents_Dynamic}
\begin{tabular}{M{1.5cm}M{1.4cm}|M{3cm}M{1cm}|M{1cm}M{1cm}|M{1cm}M{1cm}|M{1cm}M{2cm}}
\toprule

\multicolumn{3}{c}{\multirow{2}{*}{\textbf{Data Name}}} & 
\multicolumn{6}{c}{\textbf{Populated \%}} & 
\multirow{2}{*}{\textbf{Sampling Time}} \\
\cmidrule{4-9}

\multicolumn{3}{c}{} & \multicolumn{2}{c}{\textbf{ICE Vehicle \& HEV}} & \multicolumn{2}{c}{\textbf{PHEV}} & 
\multicolumn{2}{c}{\textbf{EV}} & \\
\toprule

\textbf{GPS} & \multicolumn{2}{c}{Latitude / Longitude (deg)} & \multicolumn{2}{c}{100 \%} & \multicolumn{2}{c}{100 \%} & \multicolumn{2}{c}{100 \%} & 3 sec \\
\midrule

\multicolumn{3}{c}{Vehicle Speed (km/h)} & \multicolumn{2}{c}{100 \%} & \multicolumn{2}{c}{100 \%} & \multicolumn{2}{c}{100 \%} & 1 sec \\
\midrule

\multirow{8}{*}{\parbox{1cm}{\textbf{Engine \\ Info}}} &
\multicolumn{2}{c}{Engine RPM (rev/min)} & \multicolumn{2}{c}{99.98 \%} & \multicolumn{2}{c}{100 \%} & \multicolumn{2}{c}{0 \%} & 2 sec \\
\cmidrule{2-10}

& \multirow{3}{*}{\parbox{0.8cm}{\textbf{Fuel \\ Info}}} & Mass Air Flow (g/s) & \multirow{3}{*}{100 \%} & 80 \% & \multirow{3}{*}{100 \%} & 50 \% &  \multirow{3}{*}{0 \%} & 0 \% & 2 sec \\
& & Fuel Rate (L/h) & & 0.3 \% & & 50 \% & & 0 \% & 5 sec \\
& & Absolute Load (\%) & & 83.9 \% & & 0 \% & & 0 \% & 5 sec \\
\cmidrule{2-10}

& \multicolumn{2}{c}{Short Term Fuel Trim B1 (\%)} & \multicolumn{2}{c}{89.2 \%} & \multicolumn{2}{c}{15.1 \%} & \multicolumn{2}{c}{0 \%} & 5 sec \\
& \multicolumn{2}{c}{Short Term Fuel Trim B2 (\%)} & \multicolumn{2}{c}{33.0 \%} & \multicolumn{2}{c}{0 \%} & \multicolumn{2}{c}{0 \%} & 5 sec \\
& \multicolumn{2}{c}{Long Term Fuel Trim B1 (\%)} & \multicolumn{2}{c}{79.8 \%} & \multicolumn{2}{c}{15.1 \%} & \multicolumn{2}{c}{0 \%} & 30 sec \\
& \multicolumn{2}{c}{Long Term Fuel Trim B2 (\%)} & \multicolumn{2}{c}{23.8 \%} & \multicolumn{2}{c}{0 \%} & \multicolumn{2}{c}{0 \%} & 30 sec \\
\midrule

\multicolumn{3}{c}{Outside Air Temperature ($^\circ$C)} & \multicolumn{2}{c}{43.6 \%} & \multicolumn{2}{c}{100 \%} & \multicolumn{2}{c}{100 \%} & 60 sec \\
\midrule

\multirow{3}{*}{\parbox{1cm}{\textbf{Auxiliary Power (HVAC)}}} & \multicolumn{2}{c}{AirCon Power (KW)} & \multirow{2}{*}{0 \%} & 0 \% & \multirow{2}{*}{100 \%} & 50 \% & \multirow{2}{*}{100 \%} & 100 \% & \multirow{3}{*}{60 sec} \\
& \multicolumn{2}{c}{AirCon Power (W)} & & 0 \% & & 50 \% & & 0 \% &  \\
& \multicolumn{2}{c}{Heater Power (W)} & \multicolumn{2}{c}{0 \%} & \multicolumn{2}{c}{19.4 \%} & \multicolumn{2}{c}{100 \%} &  \\
\midrule

\multirow{3}{*}{\parbox{1cm}{\textbf{Battery Info}}} & 
\multicolumn{2}{c}{Battery SOC (\%)} & \multicolumn{2}{c}{0 \%} & \multicolumn{2}{c}{100 \%} & \multicolumn{2}{c}{100 \%} & 60 sec \\
& \multicolumn{2}{c}{Battery Voltage (V)} & \multicolumn{2}{c}{0 \%} & \multicolumn{2}{c}{100 \%} & \multicolumn{2}{c}{100 \%} & 5 sec \\
& \multicolumn{2}{c}{Battery Current (A)} & \multicolumn{2}{c}{0 \%} & \multicolumn{2}{c}{100 \%} & \multicolumn{2}{c}{100 \%} & 1 sec \\

\bottomrule
\end{tabular}
\end{table*}

The dataset consists of two primary components: (1) static data and (2) dynamic time-series data. Static data include vehicle parameters described in Table.~\ref{Table:Data_Contents_Static}. Dynamic data include time-stamped naturalistic driving records of the fleet depicted in Table.~\ref{Table:Data_Contents_Dynamic}.

\begin{table}[h]
\caption{Contents of Static Data}
\label{Table:Data_Contents_Static}
\begin{tabular}{
>{\centering\arraybackslash}m{3cm} 
>{\centering\arraybackslash}m{5cm} }
\toprule
\textbf{Parameter} & \textbf{Example Values}  \\
\toprule
Vehicle Type & ICE Vehicle, HEV, PHEV, EV \\
\midrule
Vehicle Class & Passenger Car, SUV, Light truck \\
\midrule
Engine Configuration & I4, V4, V4 Flex, V6 PZEV \\
\midrule
Engine Displacement & 1.0L, 2.0L, 3.6L \\
\midrule
Transmission & 5-SP Automatic, 4-SP Manual, CVT \\
\midrule
Drive Wheels & FWD, AWD \\
\midrule
Vehicle Weights & 3,000lb, 5,000lb \\
\bottomrule
\end{tabular}
\end{table}

The list of dynamic signals, described in Table.~\ref{Table:Data_Contents_Dynamic}, are categorized into the three groups: GPS signals, standard OBD-II signals, and OEM-customized OBD-II signals. \\

\noindent \textbf{GPS Signals}.
The \textit{latitude}, and \textit{longitude} are reported by the OBD-II data loggers. The GPS signals report 7 digits for both \textit{latitude}, and \textit{longitude} as they were recorded as floating point numbers with 32-bit precision. While GPS altitude information are not provided, they can be obtained using commercial APIs, such as the Google Maps API. \\
	
\noindent \textbf{Standard OBD-II Signals}.
As discussed in the previous section, standard OBD-II signals include vehicle speed, engine related signals, and ambient air temperature. The details on the sources of the standard OBD-II signals as well as their resolutions are defined in \cite{ref:SAE_AbsLoad}. For example, the resolution of \textit{Vehicle Speed} is $1$ $km/h$, and it is either derived from a vehicle speed sensor, or calculated by the ECU using other sensors, or through the vehicle data bus. \\

\noindent \textbf{OEM-customized OBD-II Signals}.
Among the broad range of OEM-customized OBD-II signals, our OBD-II data loggers collected information on battery usage and auxiliary power consumption. \textit{Battery Current} and \textit{Battery Voltage} indicate DC current and voltage measured at the high voltage terminal. While \textit{Battery Voltage} can only have non-negative values, \textit{Battery Current} can be positive (charging) or negative (discharging). \textit{Battery SOC} is the state of charge of the high voltage battery as a percentage of maximum battery capacity. \textit{AirCon Power} and \textit{Heater Power} represent the power consumed by the AC compressor and the electric heater, respectively.
\\

The list of the dynamic data is divided into three groups depending on the availability of the signals; the first group with ICE vehicles and HEVs, the second group with PHEVs, and the last group with EVs. All groups reported GPS signals, however, battery and auxiliary power are not available in the ICE-HEV group, and the engine-related signals are not available in the EV group.
	
ICE vehicles run on heat engines relying upon the combustion of fuel in air. Among 264 distinct ICE vehicles in the dataset, there are 107 unique models from 26 different automobile manufacturers. 

HEVs include both an ICE and an electric motor. HEVs cannot be recharged by plugging into an outlet, instead relying upon regenerative braking, which converts kinetic energy waste into electricity or redirection of power from the engine. In the dataset, there are 92 distinct HEVs of 18 different models from 10 different automobile manufacturers.

PHEVs, like HEVs, have both an ICE and an electric motor (or two), except batteries of a PHEV can be charged by wall power. 24 PHEVs of 4 different models are included in the fleet.

EVs, or battery electric vehicles (BEVs) operate solely on electricity through on-board batteries. An EV does not have an ICE, and its electric driving range is longer than that of a PHEV. The dataset includes 3 EVs.

\subsection{Estimation of fuel consumption}

In most cases, fuel consumption (FC) is not directly reported by the OBD-II. However, FC can be calculated from engine related signals. Below is the pseudocode for the FC estimation method which is based on \cite{ref:SAE_AbsLoad} and \cite{ref:FCCorection_kolmanovsky}.

\begin{algorithm}
    \SetKwInOut{Input}{Input}
    \SetKwInOut{Output}{Output}

    \Input{$FuelRate$, $MAF$, $AbsLoad$, $Displacement_{eng}$, $RPM_{eng}$, $STFT$, $LTFT$, $AFR$, $\rho_{air}$}
    \Output{$FCR$}
    $correction$ = (1 + $STFT$/100 + $LTFT$/100)/$AFR$ \\
    \uIf{$FuelRate$ is available}
      {
        return $FuelRate$
      }
    \uElseIf{$MAF$ is available}
      {
        return $MAF$ * $correction$
      }
    \uElseIf{$AbsLoad$ and $RPM_{eng}$ are available}
      {
        $MAF$ = $AbsLoad$/100*$\rho_{air}$*$Displacement_{eng}$*$RPM_{eng}$/120 \\
        return $MAF$ * $correction$ 
      }
    \uElse
      {
        return NaN
      }
    \caption{Estimation of Fuel Consumption Rate (FCR)}
    \label{Algo:Estimation_FCR}
\end{algorithm}

$FuelRate$, $MAF$, $AbsLoad$, $Displacement_{eng}$, $RPM_{eng}$, $\rho_{air}$, $STFT$, $LTFT$, and $AFR$ indicate fuel consumption rate ($L/hr$), mass air flow ($g/s$), absolute engine load (\%), engine cylinder displacement ($L$), engine revolution per minute ($rev/min$), air density ($kg/m^{3}$), short term fuel trim (\%), long term fuel trim (\%), and air-fuel ratio.

For most ICE vehicles and HEVs, $FuelRate$, the FC estimates reported by the vehicle OEM, is not available. Thus, Algorithm.~\ref{Algo:Estimation_FCR} relies primarily on $MAF$ for the estimation of $FCR$. When $MAF$ is not available, a set of engine signals are used to estimate $MAF$ which is then used to obtain $FCR$ \cite{ref:SAE_AbsLoad}.

The ratio between air and fuel mass ($AFR$) is critical for accurate FC estimation. An engine operates under various $AFR$ values depending on the driving condition. To maintain a balanced ratio between air and fuel, the variations are adjusted through the fuel trims. Short term fuel trims refer to instantaneous adjustments and long term fuel trims refer to long-term adjustments which compensate for gradual changes that occur over time. In order to improve estimation accuracy, a correction factor should be applied. Algorithm.~\ref{Algo:Estimation_FCR} utilizes $STFT$ and $LTFT$ to compute the correction factor, $correction$ \cite{ref:FCCorection_kolmanovsky}. 
Since the dynamic $AFR$ is not available in the dataset, one can use the stoichiometric ratio (14.08 for E10 fuel) for $AFR$. For vehicles with $STFT$ and/or $LTFT$ of bank 2, the correction factor was calculated by taking the average of $STFT$ and/or $LTFT$ of the two banks. Other methods to reduce the noise inherent in the signals can also be applied.

Instantaneous electric energy consumption rate of PHEV \& EV can simply be obtained from multiplication of instantaneous battery voltage and current. Battery SOC and filtering can be supplemented to improve the accuracy of the energy consumption estimates.

\section{De-identification}

Signals in the dataset are not unique identifiers for individual vehicles; however, the start and/or the end of GPS coordinates with known vehicle make and model may reveal private information about the owner of the vehicle, including where he or she lives, specific locations of the driver at specific times, and personal points of interest.

To address such privacy concerns, the raw data is de-identified, or anonymized, prior to its public consumption. While there are powerful anonymization methods to sanitize static geographic datasets including methods which guarantee the privacy in terms of $k$-anonymity \cite{ref:DI_offline_kAnonymity}, \cite{ref:DI_survey_random_perturbation}, there is a trade-off between the resolution of data and the privacy. The methods which greatly sacrifice spatio-temporal resolution of data are not suitable for our purpose in which the roads and routes that a vehicle traveled should be uniquely determined by its GPS traces.

We developed a de-identification method to protect privacy of individual drivers while preserving the resolution of data and crucial content. Our method focuses on suppression and generalization of GPS trajectories of individual drivers. The method is threefold: (1) random masking of GPS coordinates of the trip start and the trip end (random-fogging), (2) geographic fencing of the trips (geo-fencing), and (3) bounding trips between major intersections (major intersections bounding). In addition to the de-identification on GPS trajectories, vehicle specific information including maker, model, year, trim level, weight, and other details are either suppressed or generalized. However, general powertrain information such as transmission and engine type are preserved.

\subsection{Random-fogging}
\begin{figure}[h] 
    \centering
    \includegraphics[width=0.9\linewidth]{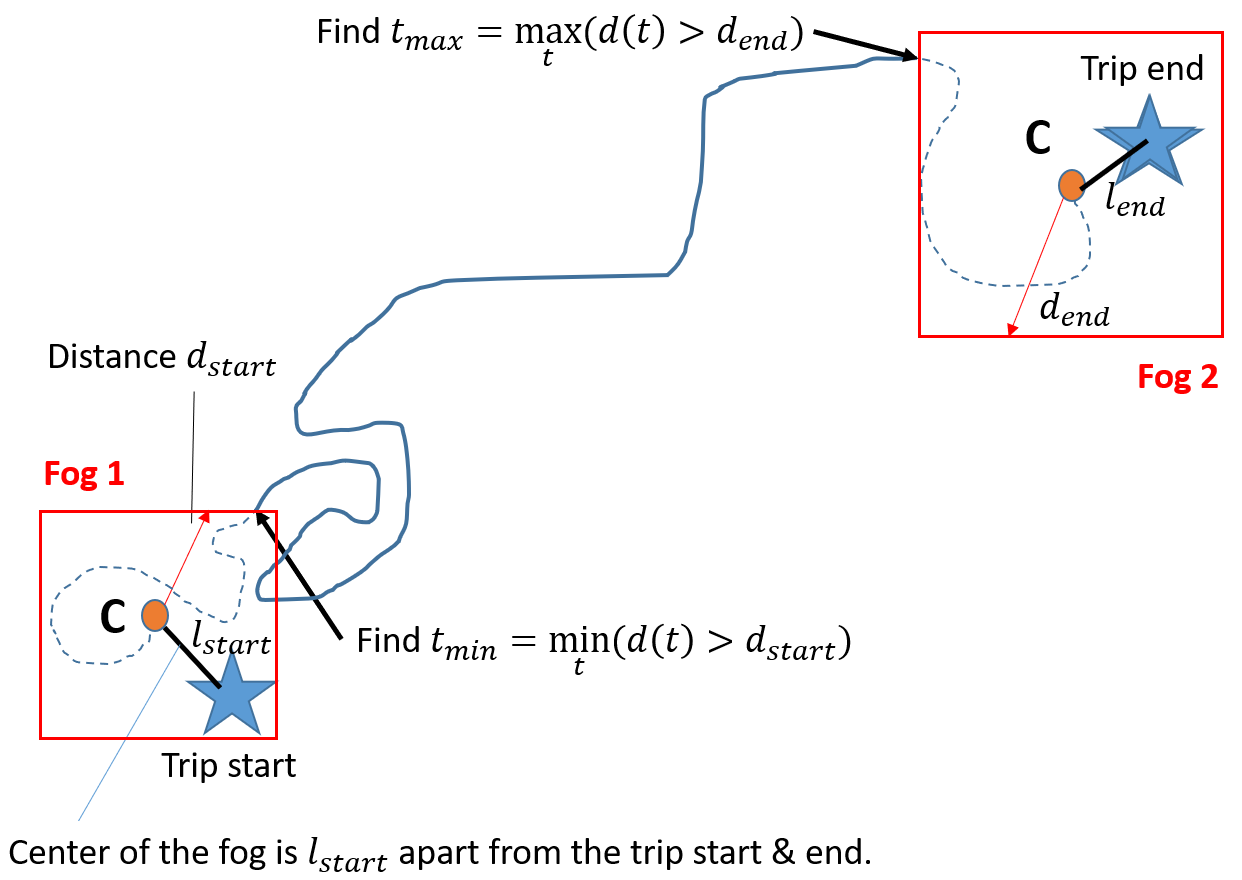}
    \caption{Random portions of a trip at the start and the end are removed as a mean of spatio-temporal suppression.}
    \label{Fig:Random_Fog}
\end{figure}

The first component of our de-identification method is to randomly mask GPS trajectories of the start and the end of a trip. We first generate two objects ("fogs") whose boundaries are parametric equations of latitude and longitude. Next, two fogs are placed near the start and the end of trips, so that the center of the fogs are $l_{start}$ and $l_{end}$ apart from the trip start and the trip end.  The minimum time $t_{min}$ and $t_{max}$ are obtained from the equation shown in Fig.~\ref{Fig:Random_Fog} based on random $d_{start}$ and $d_{end}$. Lastly, the GPS trajectories recorded prior to $t_{min}$ and after $t_{max}$ are suppressed. The parameters of the fogs, the distance metric, as well as the number of sub-fogs within a fog are chosen at random.

%% To do: Maybe writing a pseudocode for the random-fog algorithm?

\subsection{Geo-fencing}
The second component of the de-identification method is fencing recordings within a geographic boundary. The raw dataset includes trips conducted in broad south-east Michigan area where density of the roads greatly vary from urban to rural settings. The density affects the effectiveness of a de-identification method, as it can be associated with a number of unique routes a driver can take.

\begin{figure}[h] 
    \centering
    \includegraphics[width=\linewidth]{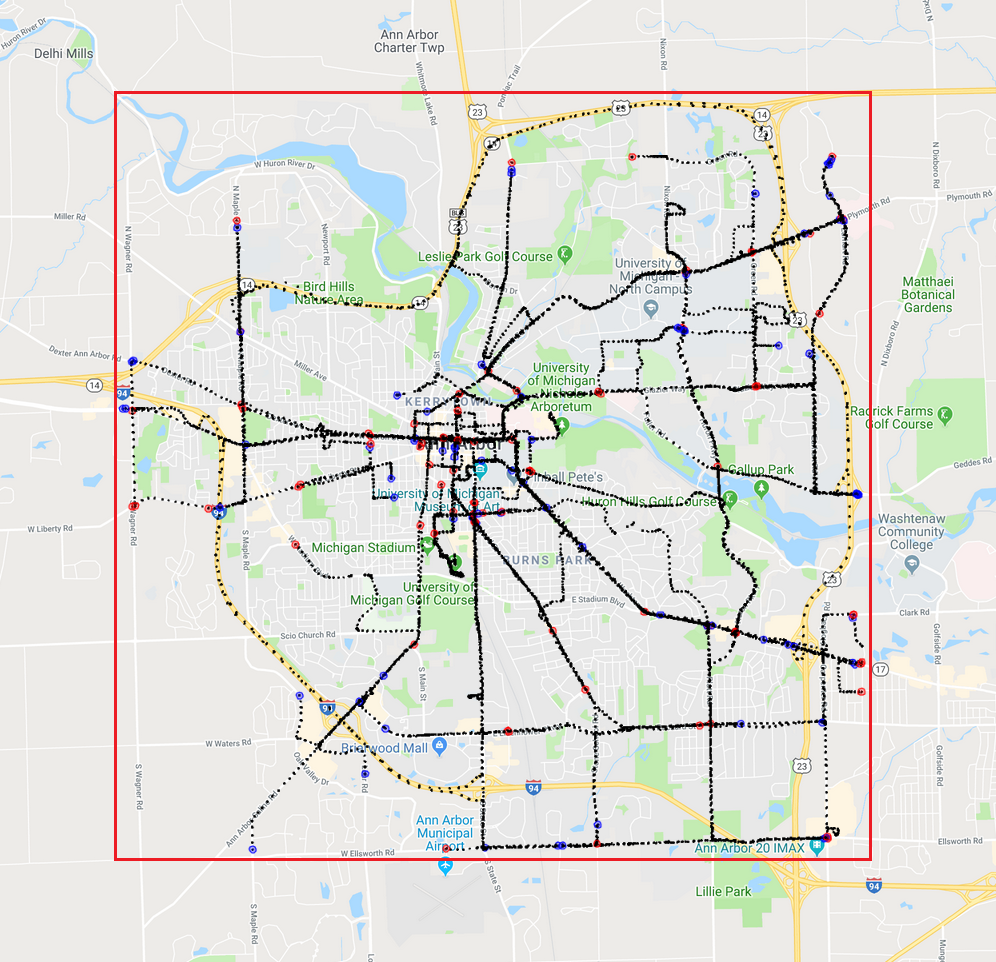}
    \caption{Geo-fencing described as the red box and major intersections bounding are applied. The blue and red dots represent the new trip start (the first major intersection) and the new trip end (the last major intersection). The portion of a trip prior to the new trip start and subsequent to the new trip end are removed.}
    \label{Fig:GeoFence_MNodes}
\end{figure}

In this sense, we only preserved the data within the city of Ann Arbor where the density of roads is high by applying a rectangular geometric fence and removing the data outside the fence as described in Fig.~\ref{Fig:GeoFence_MNodes}. This removes GPS trajectories in rural areas where houses are sparse, and the number of unique paths are low, which would facilitate re-identification. This also serves as fail-safe measure of random-fog and major intersection bounding.

\subsection{Major Intersections Bounding}
The majority of the personally identifiable information inherent in the raw GPS trajectories are from the start and the end of the trips, as they can be used to identify the locations of participants' homes and workplaces. The third component of the anonymization method is to suppress data near the trip start and the trip end. This is done by only preserving GPS trajectories in between the first and the last major intersections through which a vehicle traveled as described in Fig.~\ref{Fig:GeoFence_MNodes}.

The parameters used in the random fog, and the order of the three methods are not discussed to better protect participants' privacy.

\section{Driver Behaviors on Fuel Consumption}
In this section, we demonstrate how VED can be utilized for fuel consumption behaviors of human drivers. We first discuss the influence of driving environments on fuel consumption and vehicle speed. Next, the relationship between fuel economy and vehicle speed is investigated. Third, we demonstrate the impacts of seasons on miles per gallon (MPG) and speed of vehicles. The relationship between MPG and ambient temperature is identified. Lastly, we examine how time of day (TOD) affects MPG and travel speed.

In order to obtain heatmaps for Fig.~\ref{Fig:DrivingEnv_All}, \ref{Fig:Season_Impact_I94}, \ref{Fig:TOD_RushHour}-\ref{Fig:Plymouth_ICEvsHybrid}, we first post-processed the data to select the area of interest and discretized the roads into a number of grids. Then, we calculated median and mean values of the signals of interest.

\subsection{Influence of driving environments on Fuel Economy}

\begin{figure*}[ht] 
    \centering
    \includegraphics[width=\linewidth]{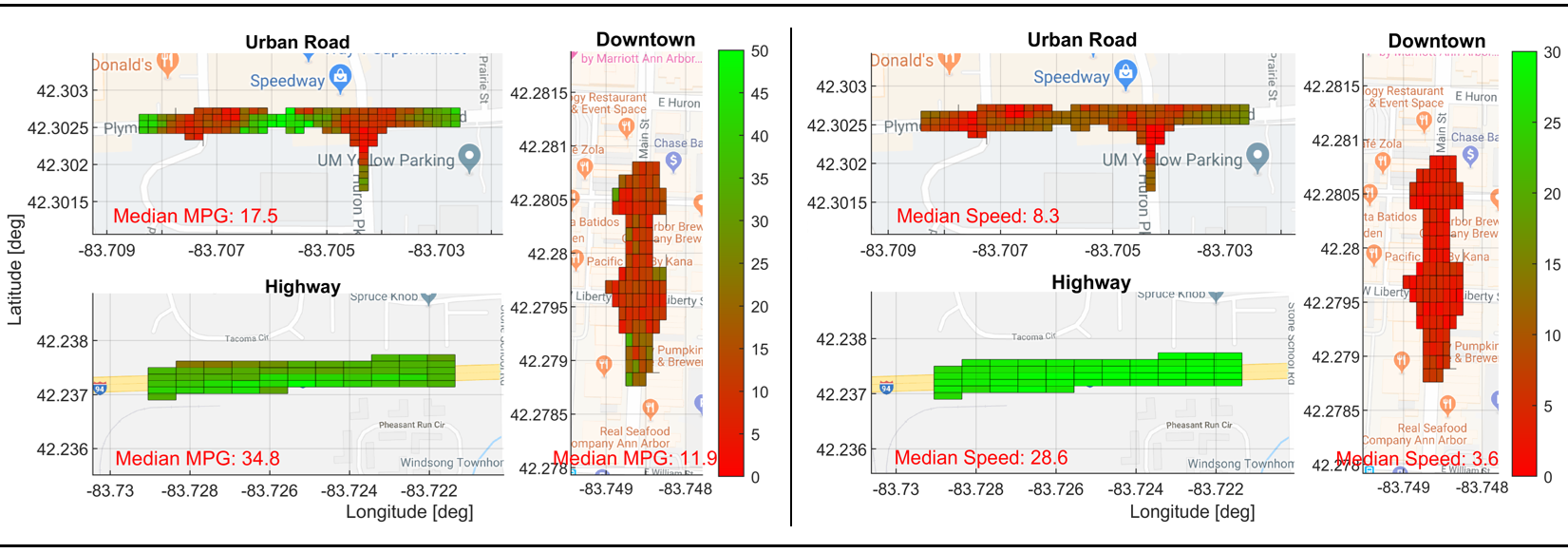}
    \caption{The three plots on the left represent MPG heatmaps conducted on the three driving environments. The number of trips and vehicles traveled on each environment are 2343/232, 675/150, 656/159 for the urban road, the highway, and the downtown. The three plots on the right represent travel speed heatmaps of the same trips. The median MPGs and speeds are described in the figure. The posted speed limits of the driving environments are 15.6 $m/s$ (35mph), 31.3 $m/s$ (70mph), 11.2 $m/s$ (25mph) for the urban roads, the highways, and the downtown roads.}
    \label{Fig:DrivingEnv_All}
\end{figure*}

Daily driving occurs in various driving environments including urban roads and highways. On urban roads, drivers are exposed to dense traffic and traffic signals, resulting frequent speed fluctuations. On the other hand, on highways, the speed is higher and the traffic density is relatively low, thus allowing drivers to maintain constant travel speed. Such behaviors have a significant influence on fuel consumption.

In this subsection, we investigate the impact of driving environments on fuel consumption and travel speed of all ICEs and HEVs with fuel signals. In particular, we demonstrate the following: (1) Impact of driving environments on MPG and travel speed; and (2) Fuel consumption heatmap and identification of fuel saving opportunities in urban driving environments.

The first environment we studied is a section of Plymouth road in North Ann Arbor, which has two lanes in each direction and two consecutive signalized intersections, Plymouth-Nixon and Plymouth-Huron Parkway. It represents the busy urban roads with closely-spaced traffic lights. 

The second driving environment is a section of Interstate 94 (I-94) in South Ann Arbor, which consists of two lanes in each direction. This section represents the highway driving environments.

The third driving environment is a single-lane road in downtown Ann Arbor. This environment includes two of the busiest signalized intersections, Main-Washington and Main-Liberty, in the city of Ann Arbor. This section represents business districts with heavy vehicle and pedestrian traffic.

%%%%%%%%%%%%%%%%%%%%%%%%%%%%%%%%%%%%%%%%%%%%%%%%%%%%555
\begin{comment}
\begin{figure}[h] 
    \centering
    \includegraphics[width=\linewidth]{Figures/Plymouth_All.png}
    \caption{MPG and travel speed of the vehicles on Plymouth road, North Ann Arbor. The median MPG and speed is 17.5 miles per gallon and 8.3 meters per seconds. The speed limit is 15.6$m/s$, or 35mph.}
    \label{Fig:Plymouth}
\end{figure}
\begin{figure}[h] 
    \centering
    \includegraphics[width=\linewidth]{Figures/I94_All.png}
    \caption{MPG and travel speed of the vehicles on I-94, in South Ann Arbor. The median MPG and speed is 34.8 miles per gallon and 28.6 meters per seconds. The speed limit is 31.3$m/s$, or 70mph.}
    \label{Fig:I-94}
\end{figure}
\begin{figure}[h] 
    \centering
    \includegraphics[width=\linewidth]{Figures/Downtown_All.png}
    \caption{MPG and travel speed of the vehicles on Main street, in downtown Ann Arbor. The median MPG and speed is 11.9 miles per gallon and 3.6 meters per seconds. The speed limit is 11.2$m/s$, or 25mph.}
    \label{Fig:Downtown}
\end{figure}
\end{comment}
%%%%%%%%%%%%%%%%%%%%%%%%%%%%%%%%%%%%%%%%%%%%%%%%%%%%555

The results show that both MPG and travel speed are the highest on the highway, and the lowest in downtown Ann Arbor. While these results align with our common knowledge of fuel consumption in urban roads and highways, it was observed that MPG on the highway was almost twice as high as that on urban roads. 

Heatmaps in Fig.~\ref{Fig:DrivingEnv_All} show which parts of the roads have high and low MPGs, thus identifying fuel saving opportunities. It is observed that vehicles have lower MPG near the intersections where drivers often make sudden and frequent stops, and higher MPG in the middle of roads where vehicles tend to coast and brake lightly. The results agree with a study conducted in the urban arterial roads in Harbin, China which reported that 50.36\% of the fuel was spent near intersections whose area only covers 28.9\% of the total distance traveled \cite{ref:EAD_SI_FC_China}. Indeed, the results motivate a number of eco-driving techniques including eco-driving at signalized intersections, which aims to reduce the fuel consumption by minimizing hard braking and acceleration.

One can deduce that MPG is closely related to travel speed from Fig.~\ref{Fig:DrivingEnv_All}. We studied all the 309,171 miles of data covered by 249 ICE vehicles and 90 HEVs to infer the relationship between MPG and travel speed. 

\begin{figure}[h] 
    \centering
    \includegraphics[width=0.95\linewidth]{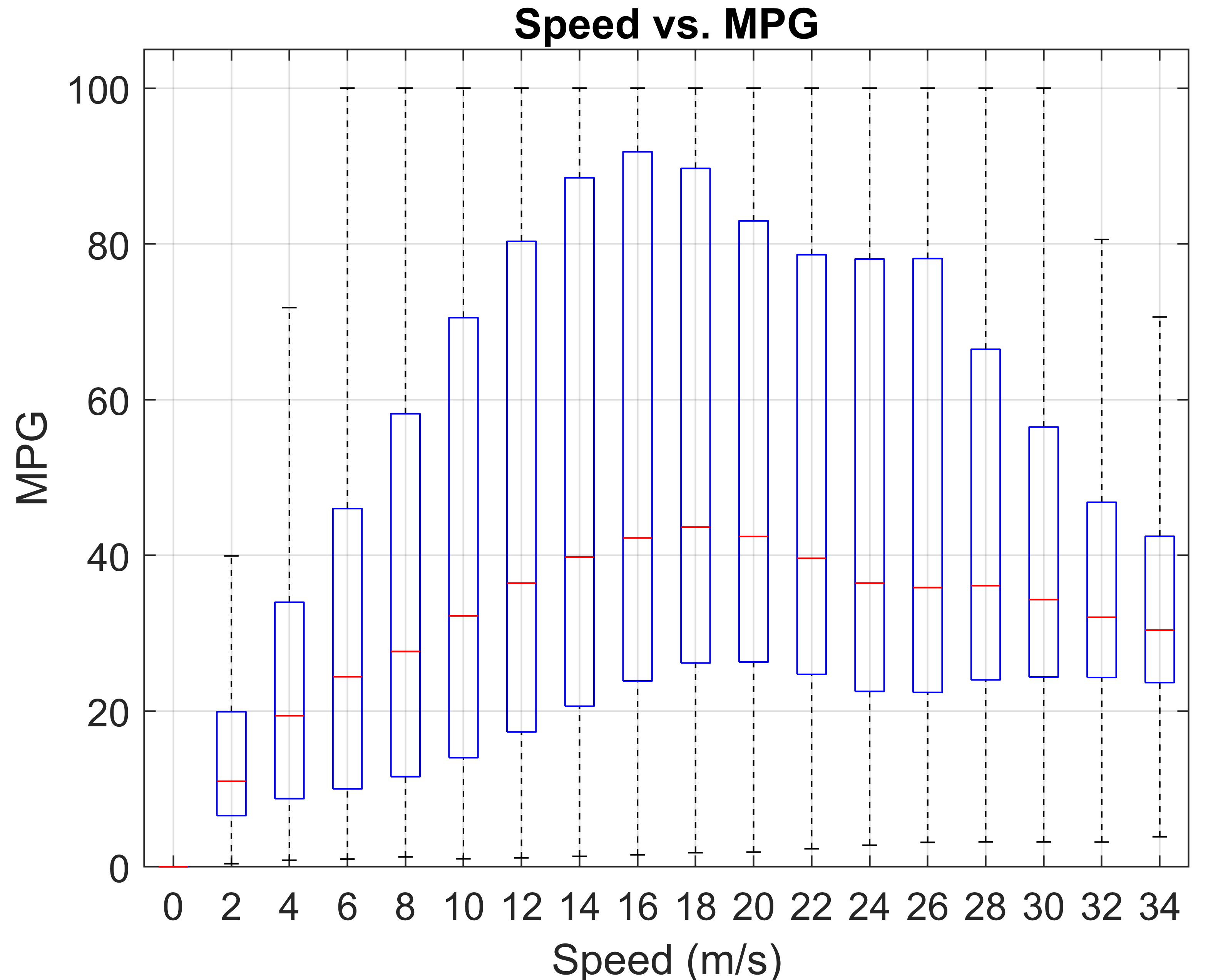}
    \caption{Box plots of MPG for all the 309,171 miles traveled by 249 ICEs and 90 HEVs in diverse real-world driving conditions from Nov, 2017 to Nov, 2018. The minimum and the maximum average MPG are 0 at 0$(m/s)$ and 43.7 at 18$(m/s)$.}
    \label{Fig:SPD_MPG}
\end{figure}

Fig.~\ref{Fig:SPD_MPG} shows the box plots which were obtained by grouping MPG into discrete bins of travel speed. The median is represented by the red band, and the top and bottom edges of the box are the first and third quartiles. Since the magnitude of MPG can be infinite for vehicles which shut engines off during coasting, an upper bound of 100 was applied.

Fig.~\ref{Fig:SPD_MPG} presents that vehicles traveling at low speed have lower MPG compared to the vehicles operating at medium and high speeds. This figure elaborates on the findings from Fig.~\ref{Fig:DrivingEnv_All}, providing one of the reasons why vehicles had higher MPG on the highway compared to urban road and downtown where the speeds are much lower. 

It is worth noting that since every vehicle has different engine and powertrain configurations, fuel consumption of two different vehicles at the same engine operating points can differ. This added to the variance of MPG, contributing to the deviations from the median value. 

The trend of MPG with respect to vehicle speed is not monotonic. In our studies, MPG peaked around 16-20$(m/s)$ and decreased with higher speed. This non-monotonicity of MPG is corroborated by other studies \cite{ref:FCvsSpd_OakRidge}, \cite{ref:FCvsSpd_FHWA1}, and \cite{ref:FCvsSpd_FHWA2}. Two studies presented by the Federal Highway Administration (FHWA) \cite{ref:FCvsSpd_FHWA1}, \cite{ref:FCvsSpd_FHWA2} showed that maximum fuel efficiency of ICE vehicles were achieved at speeds of 35-45mph, which correspond to 15.6-20.1 $m/s$, with diminished fuel efficiency at higher speeds.

\subsection{Influence of Season and Temperature on Fuel Economy}
\begin{figure}[h] 
    \centering
    \includegraphics[width=\linewidth]{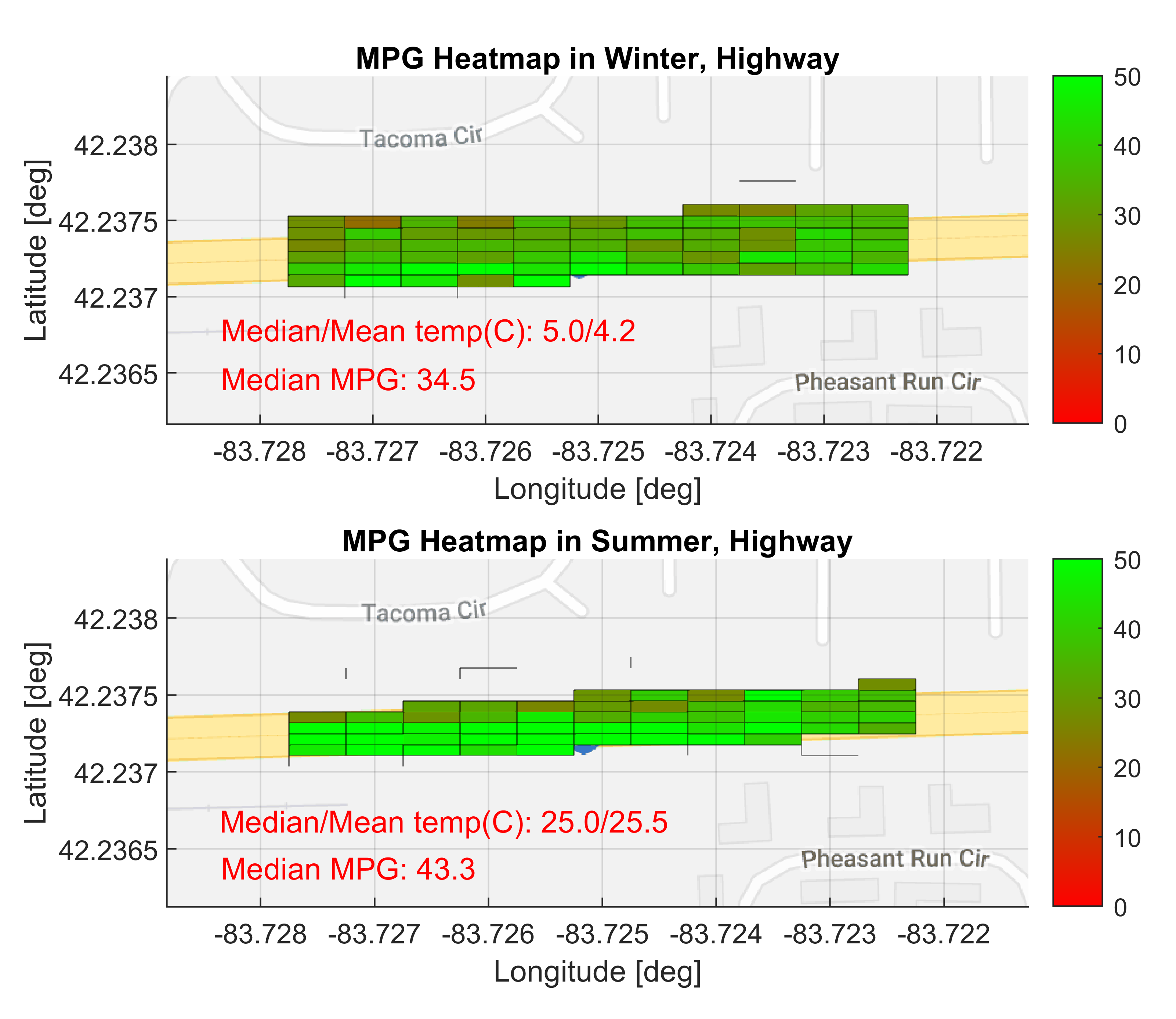}
    \caption{The top plot represents winter MPG heatmap for 160 trips of 79 vehicles that traveled through I-94 from Dec, 2017 to Jan, 2018. The bottom plot represents summer MPG heatmap for 129 trips of 46 vehicles that traveled through the same section of I-94 from Jun, 2018 to Aug, 2018. 
    The median travel speed was 28.7 $m/s$ in winter and 29.2 $m/s$ in summer. Median MPG and median/mean ambient temperatures are given in the plots.}
    \label{Fig:Season_Impact_I94}
\end{figure}

As widely reported in the literature, fuel economy is related to ambient temperature, resulting in different consumption during different seasons. In order to study these effects, we analyzed 249 ICE vehicles and 90 HEVs with valid fuel signals, and compared the results between summer and winter as shown in Fig.~\ref{Fig:Season_Impact_I94}. For this analysis, we concentrate on the section of highway I-94 in South Ann Arbor.

We observe that MPG in summer is 25.5\% higher than in winter. Due to the diversity of fleet vehicles, these results quantify the overall impact of season on fuel economy of ICE vehicles and HEVs. It should be noted that individual vehicle behavior can be vastly different.  

The above results agree with those of other
studies \cite{ref:BSFC_temp_ICE}, \cite{ref:BSFC_temp_HEV}, \cite{ref:ICEvsHEV_FC}, \cite{ref:BSFC_temp_ICE2}, \cite{ref:ICEvsHEV_FC_Zahabi2014}. It was reported that an ICE vehicle consumed 20\% more fuel over the UDDS cycle at ambient temperature of $-7^\circ$C than at $20^\circ$C \cite{ref:BSFC_temp_ICE}. An on-road evaluation was conducted to study how ambient temperatures affect the fuel economy of HEVs and reported that hybrid cars consumed more than double the fuel at $-14^\circ$C than at $25^\circ$C  \cite{ref:BSFC_temp_HEV}. A study \cite{ref:BSFC_temp_BEV} on battery electric vehicles investigated how energy efficiency varies with ambient temperature due to battery efficiency and cabin climate control. They reported energy consumption of BEVs in the Upper Midwest can increase by an average of 15\% compared to the Pacific Coast due to temperature differences.

\begin{figure}[h] 
    \centering
    \includegraphics[width=\linewidth]{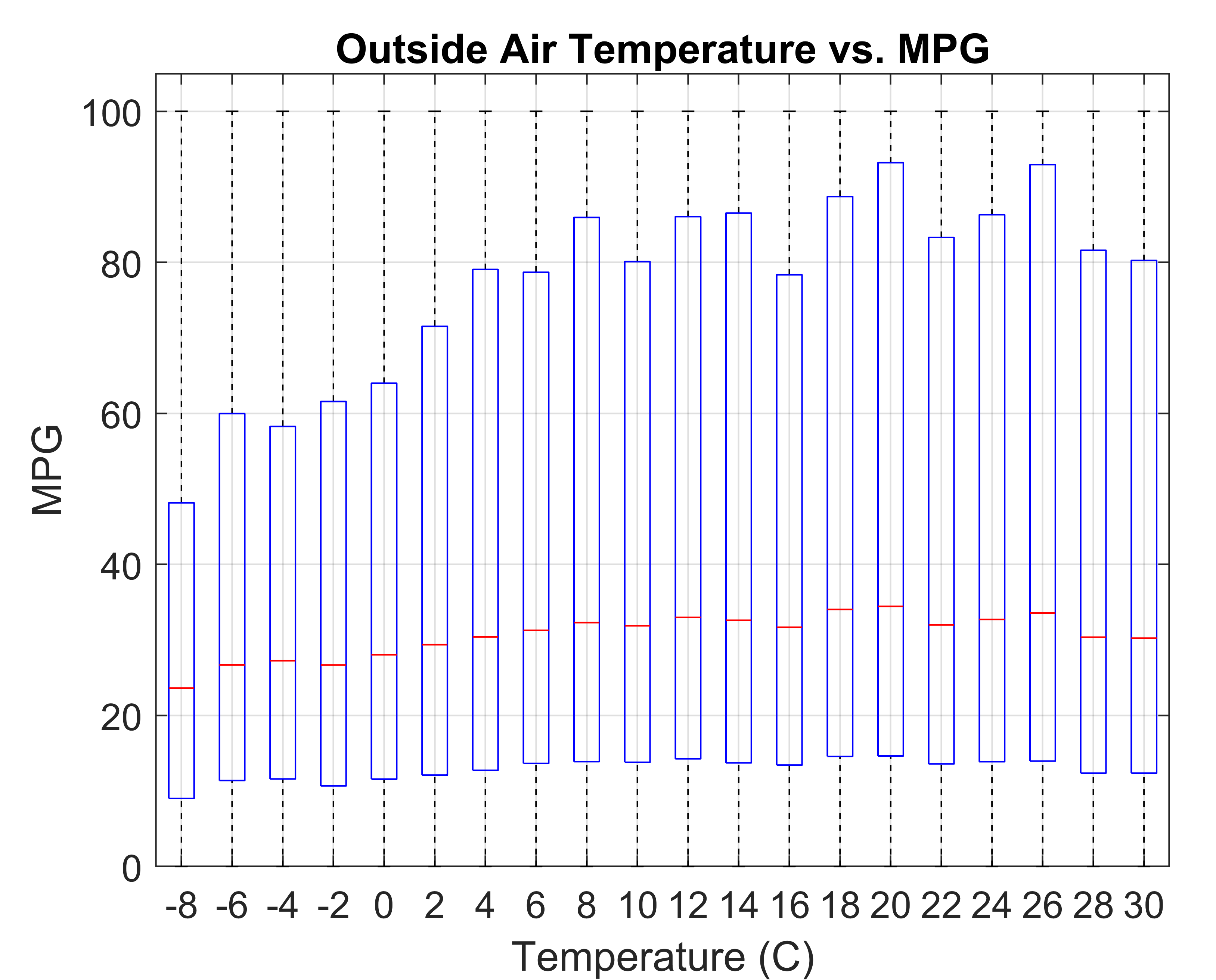}
    \caption{Box plots of MPG grouped by OAT from all vehicles with valid OAT signals. The minimum and the maximum value for median MPG are 23.6 at $-8^\circ$C and 34.4 at $20^\circ$C.}
    \label{Fig:OAT_MPG}
\end{figure}

Fig.~\ref{Fig:OAT_MPG} further elaborates on findings reported in Fig.~\ref{Fig:Season_Impact_I94}, demonstrating empirical relationship between MPG and ambient temperature, or outside air temperature (OAT). The box plots were obtained by grouping MPG by discrete bins of OAT. 

It is observed that $20^\circ$C and $-8^\circ$C were the most and the least MPG-efficient temperatures respectively, and the first and the third quartiles MPG both increased with OAT. This empirical result demonstrates the trend of fuel economy of ICE vehicles and HEVs with respect to ambient temperature.

\subsection{Influence of traffic density on MPG and speed}

It is well known that the traffic congestion level, traffic density, flow, and speed differ considerably depending on time of day (TOD). In this subsection, we quantify the influence of TOD on road traffic by analyzing speed of vehicles during rush-hours (4-6pm) and free-flow hours (9pm-6am). Furthermore, we quantify the influence of TOD on fuel economy of vehicles by analyzing MPG for the two different TOD windows.

\begin{figure}[h] 
    \centering
    \includegraphics[width=\linewidth]{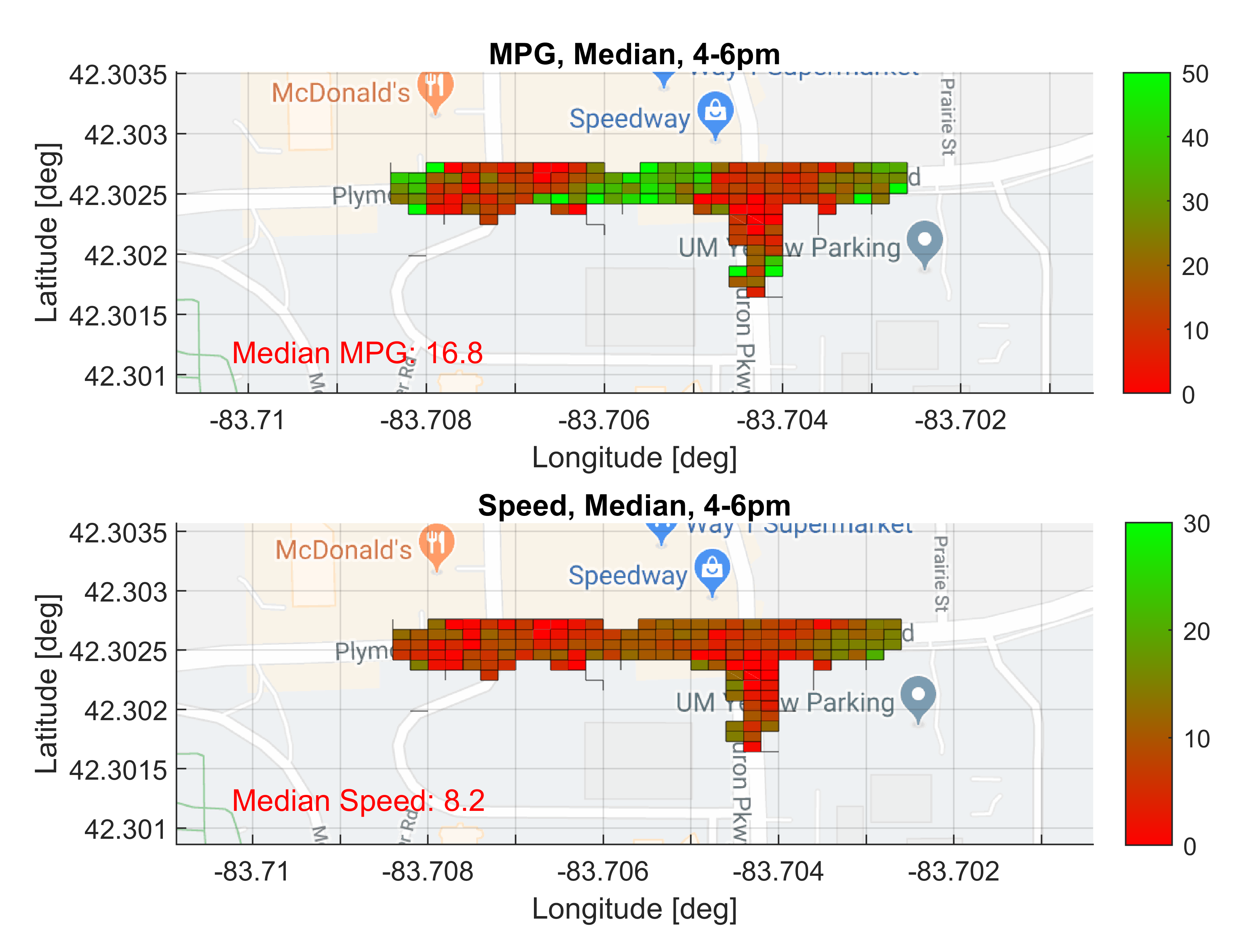}
    \caption{MPG and speed heatmaps for 378 trips of 113 ICEs and HEVs which run on the urban road described in Fig.~\ref{Fig:DrivingEnv_All} between 4pm and 6pm over the year. This TOD represents rush-hours since it's the busiest time of the day. The median MPG and speed is 16.8 and 8.2 $(m/s)$.}
    \label{Fig:TOD_RushHour}
\end{figure}

\begin{figure}[h] 
    \centering
    \includegraphics[width=\linewidth]{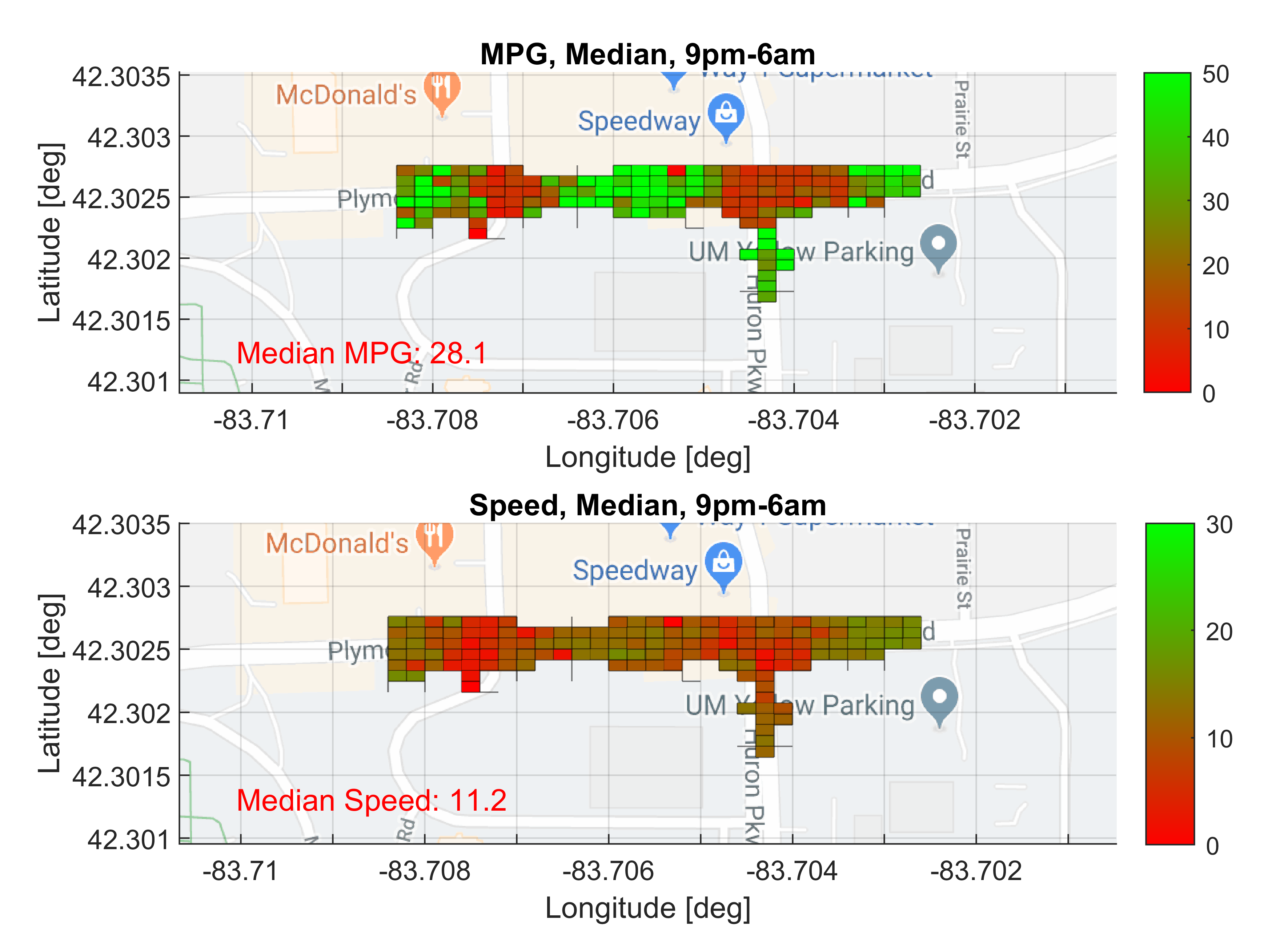}
    \caption{MPG and speed heatmaps for 156 trips of 48 ICEs and HEVs which run on the same road between 9pm and 6am over the year. This TOD represents free-flow hours since it's traffic congestion is the lowest. The median MPG and speed is 28.1 and 11.2 $(m/s)$.}
    \label{Fig:TOD_FreeFlow}
\end{figure}

Fig.~\ref{Fig:TOD_RushHour} and \ref{Fig:TOD_FreeFlow} illustrate how fuel economy and travel speed drastically vary in rush-hours and free-flow hours. A comparison of median MPG reveals a 40.2\% difference between rush hours and free-flow hours. Similarly, a comparison of median speed reveals a 26.8\% difference. One can further analyze the dataset to obtain plots of MPG versus TOD. These statistics can be utilized for making energy-efficient and the shortest-time route choices.

\section{Fuel and Energy Use of Hybrid Vehicles}
In this section, we elaborate on two case studies to demonstrate how the dataset can be utilized for fuel and energy studies of hybrid vehicles. In the first case study, we compare the fuel economy of ICE vehicles and HEVs. In the second case study, we analyze a trip conducted by a PHEV in downtown Ann Arbor, and demonstrate how trips conducted by PHEVs and EVs can be utilized.

\subsection{Fuel Economy: ICE Vehicle versus HEV}
One of the most noticeable benefits of HEVs compared to ICEs is their higher fuel efficiency. In this subsection, we analyze the driving data of HEVs and ICE vehicles in an urban road near signalized intersections and compare their fuel economy.

\begin{figure}[h] 
    \centering
    \includegraphics[width=\linewidth]{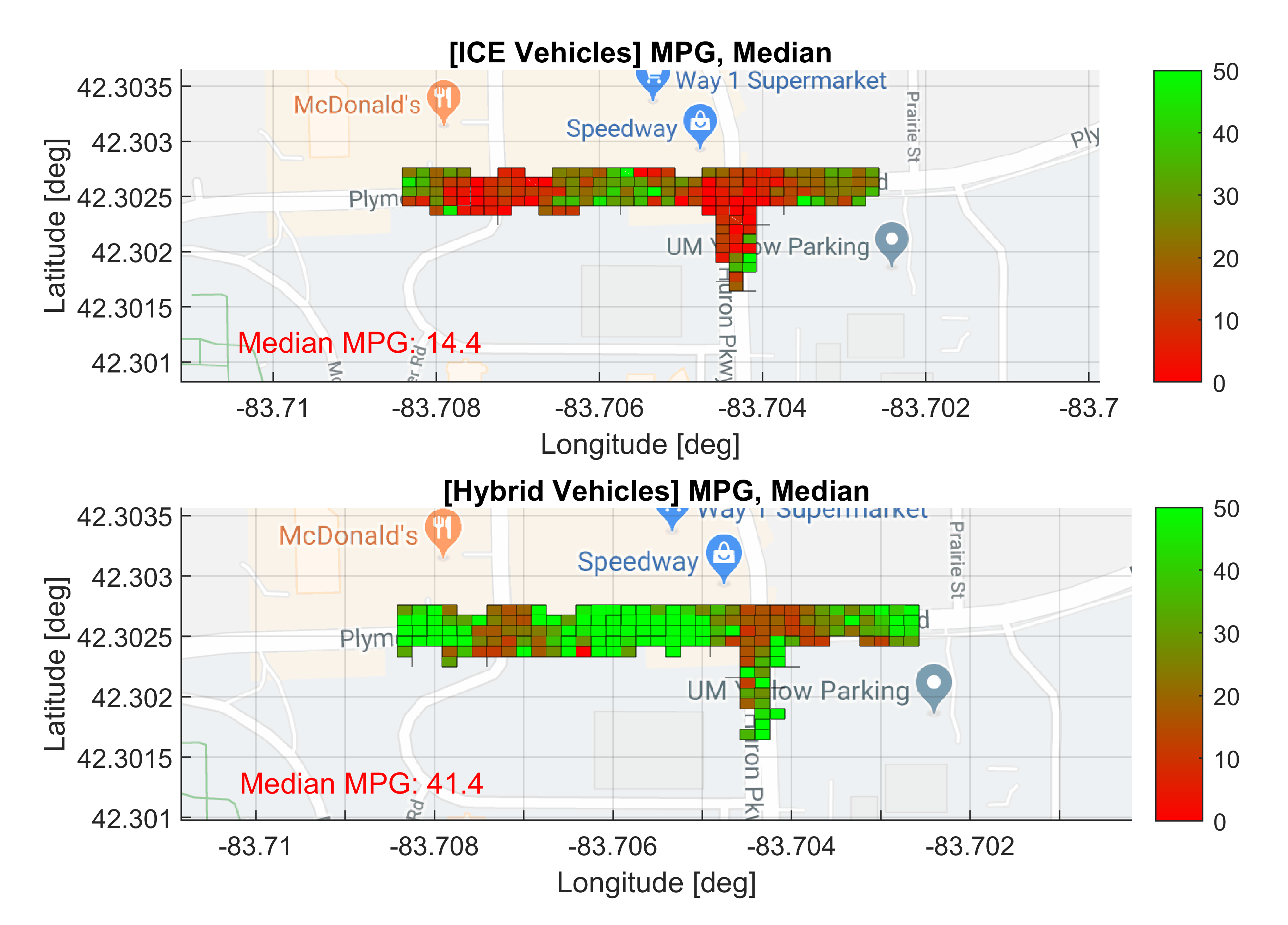}
    \caption{Top plot presents MPG heatmap for 1326 trips conducted by 158 ICE vehicles on Plymouth road. Bottom plot presents MPG heatmap for 1017 trips traveled by 74 HEVs on the same section of the road.}
    \label{Fig:Plymouth_ICEvsHybrid}
\end{figure}

Fig.~\ref{Fig:Plymouth_ICEvsHybrid} demonstrates the significant difference in fuel economy between ICEs and HEVs. It can be inferred that one of the reasons why HEVs had 2.9 times higher MPG is that they took advantage of regenerative braking, thus minimized the energy loss from severe braking near signalized intersections. 

These numbers are consistent with results from other studies. The EPA publishes annual guides on fuel economies of all types of cars including ICE vehicles, HEVs, PHEVs, and EVs \cite{ref:FEG2017_EPA}. The numbers are obtained by running vehicles through a series of EPA federal test procedures in standard conditions. A number of research \cite{ref:FC_LabtestVSRealworld}, \cite{ref:ICEvsHEV_FC_Zahabi2014}, \cite{ref:ICEvsEV_FC_China} based on real-world driving data in diverse driving conditions also showed that HEVs have higher fuel economy compared to ICE vehicles. A study \cite{ref:ICEvsHEV_FC_Zahabi2014} analyzed driving data of 21 HEVs and 53 gasoline vehicles in cold cities in Quebec, Canada, and reported that HEV sedans are 28\% more fuel efficient than gasoline sedans in cold cities. Another study \cite{ref:ICEvsEV_FC_China} examined real-world driving data of HEVs, PHEVs, and EVs in Beijing, China where the travel patterns are characterized by low speeds, long idle times, severe speed changes and short driving ranges. 

\subsection{Case Study on PHEV}
A difference between HEV and PHEV is in the operation of engines and motors. PHEVs typically operate in two distinct modes; EV mode in which the vehicles only runs on the electric motors, and Hybrid mode in which the vehicles use both engine and motor. PHEVs usually start up in EV mode and operate in this mode until a sufficient level of battery depletion or until a threshold speed is achieved. The dataset contains driving records of PHEVs in both modes.

In this subsection, we study a trip conducted by a PHEV in downtown Ann Arbor, and discuss the energy consumption of the PHEV. It is important to emphasize that HEVs did not output battery signals. One should refer to PHEV and EV data to study energy consumption of electric vehicles. 

\begin{figure}[h] 
    \centering
    \includegraphics[width=\linewidth]{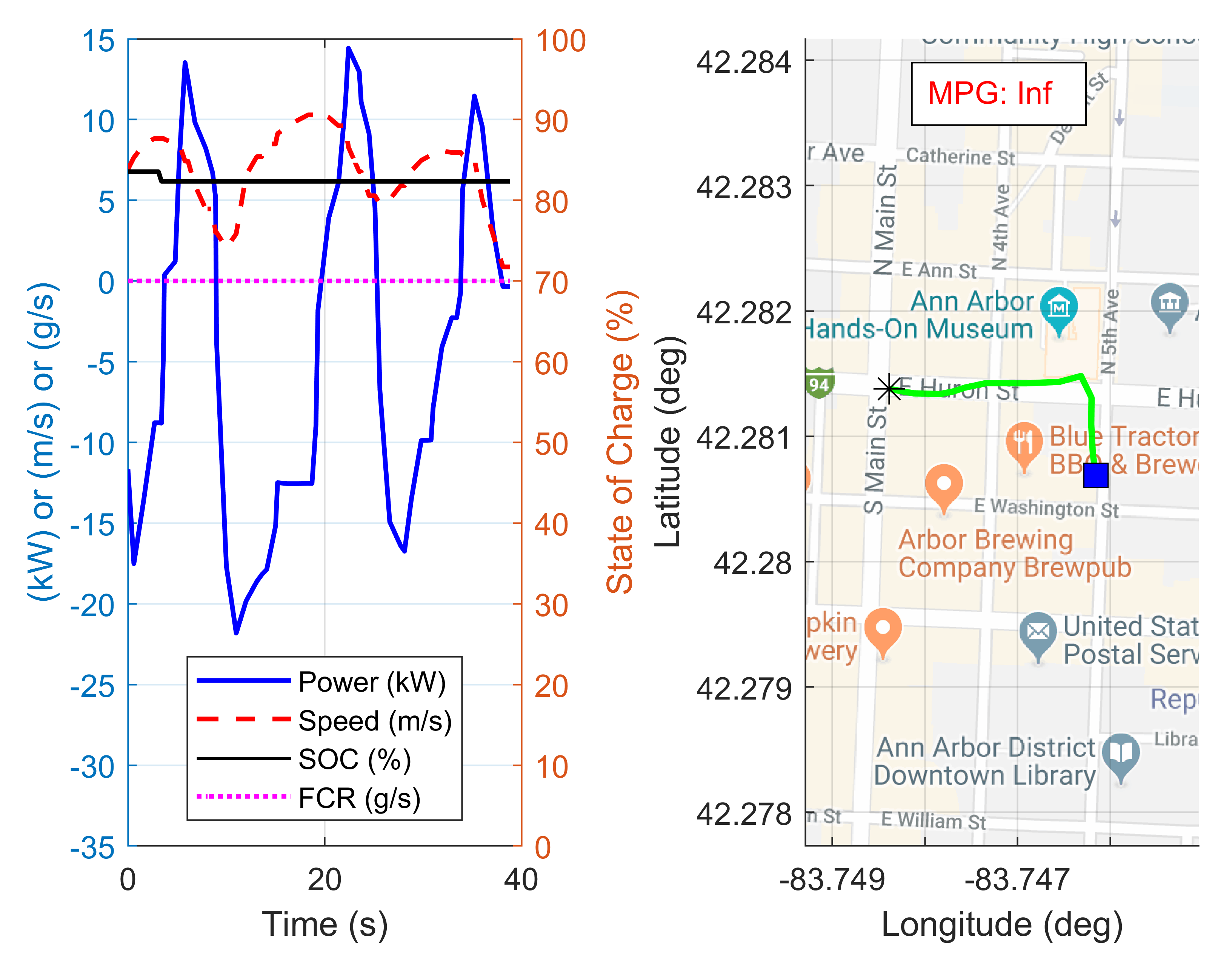}
    \caption{A trip traveled by a PHEV. The left plot presents battery power, SOC, and vehicle speed with respect to time. Positive/negative battery powers correspond to charging/depletion of the batteries. The right plot shows the corresponding path, with the x and the rectangle as the start and the end of the trip.}
    \label{Fig:PHEV}
\end{figure}

Fig.~\ref{Fig:PHEV} indicates that the vehicle operated exclusively in the EV mode, consuming no fuel, and achieving infinite MPG. The vehicle consumed 0.1265 $kWh$ traveling 0.5192$miles$, which corresponds to 138 $MPGe$, or miles per gallon equivalent, which is a new metric proposed by EPA to gauge energy efficiency of electric vehicles. This number is in line with the results published by EPA \cite{ref:FEG2017_EPA}. Another observation is that the battery was recharged during vehicle deceleration through regenerative braking, and depleted during vehicle acceleration.

\section{Eco-driving}
Eco-driving, or energy-efficient driving techniques are gaining popularity with the introduction of smart infrastructure, V2X communication, and the development of connected and automated vehicles. 

In this section, we elaborate on two applications of energy-efficient driving techniques to demonstrate how the dataset can be used for eco-driving research. We first discuss eco-routing, a routing algorithm which selectively choose routes that minimizes energy consumption. Secondly, we discuss eco-approach and departure, a planning/control algorithm which optimizes vehicle trajectories with respect to energy consumption.

\subsection{Eco-routing}
An immediate applications of the database is eco-routing. The rich fuel and energy consumption data can help users to model fuel and energy consumption of various types of vehicles on the basis of trips, routes, and links. Once fuel and energy consumption models are built, they can be utilized for route selection including energy-efficient, fastest, and shortest routes.

We demonstrate how fuel reductions can be achieved by eco-routing. We first analyze trips conducted on two different routes given the same origin-destination pair. After we obtain fuel economy statistics on the two routes, we show the difference in the MPG of the two routes. 

\begin{figure}[h] 
    \centering
    \includegraphics[width=\linewidth]{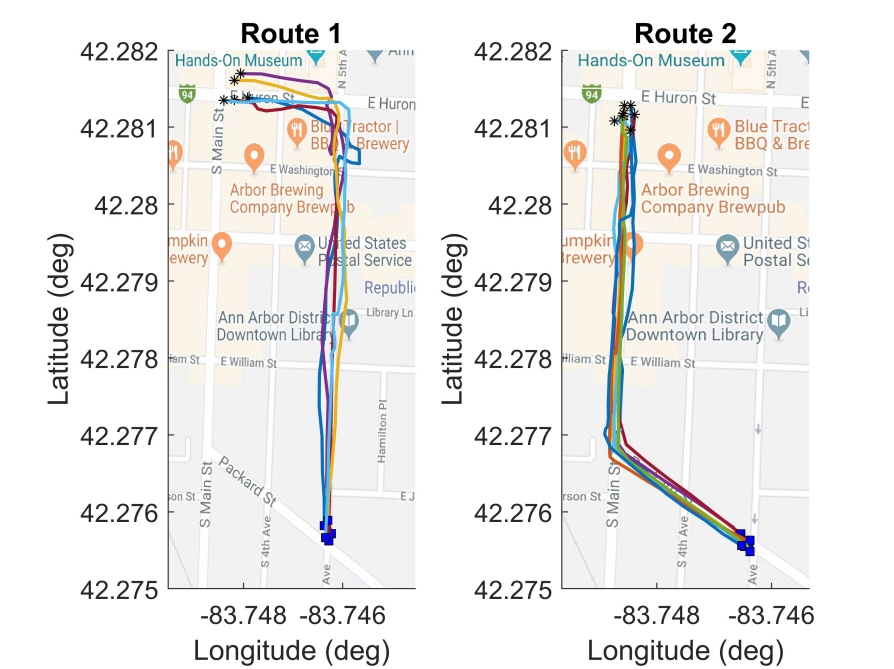}
    \caption{Trajectories of the trips conducted by ICE vehicles on the two routes, with the x and the rectangle as the start and the end of a trip.}
    \label{Fig:Eco-routing}
\end{figure}

\begin{figure}[h] 
    \centering
    \includegraphics[width=0.8\linewidth]{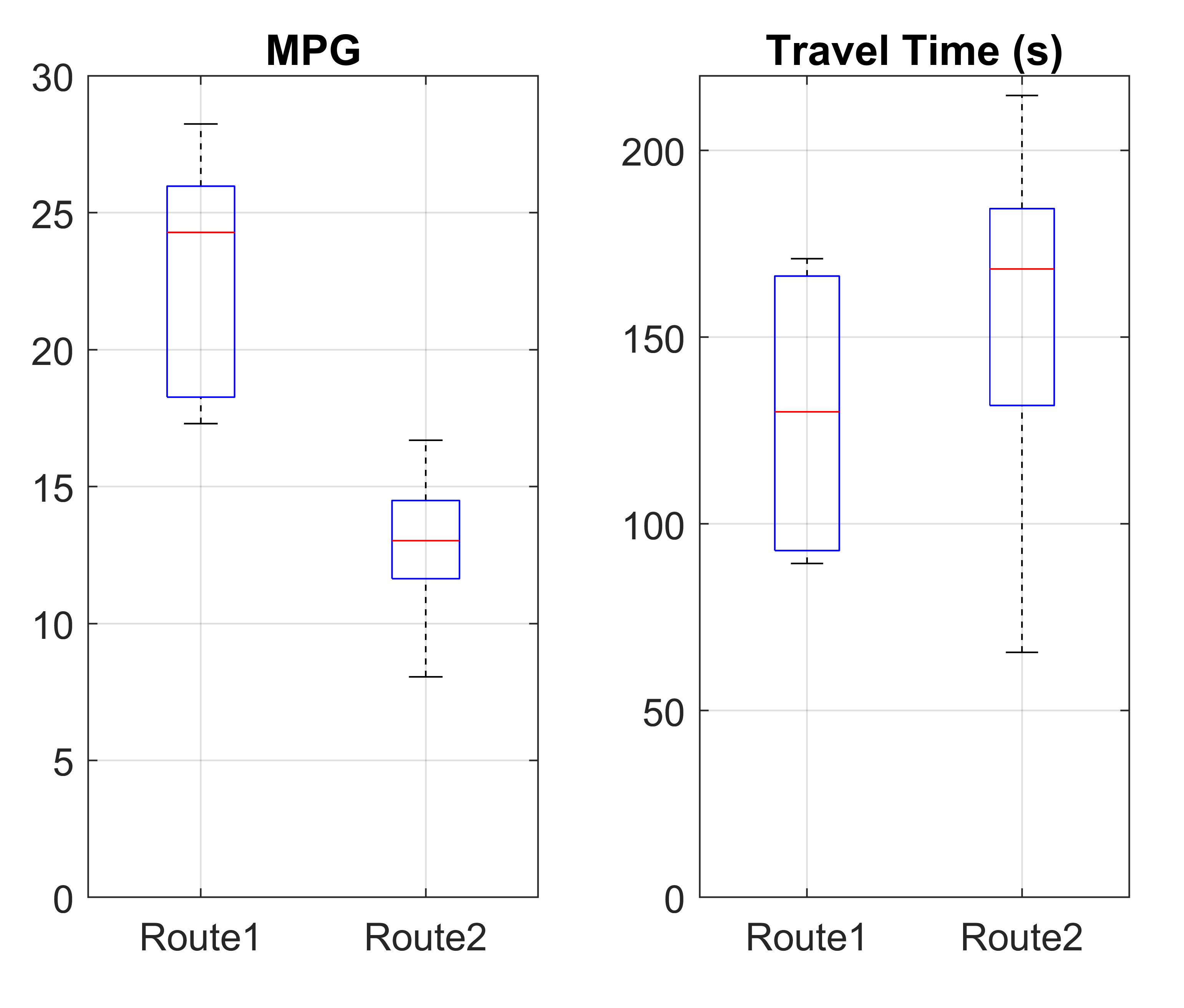}
    \caption{box plots of MPG and travel time for the trips in Fig.~\ref{Fig:Eco-routing}. The median MPG and travel time for R1/R2 are 24.3/13.0 and 130.0s/168.3s.}
    \label{Fig:Eco-routing_stats}
\end{figure}

Fig.~\ref{Fig:Eco-routing} and \ref{Fig:Eco-routing_stats} demonstrate how route choices affect the fuel economy and the travel time. The numbers indicate that there are opportunities to save fuel and travel time by choosing more fuel-efficient and/or less time-consuming routes. Other applications including data-driven fuel consumption and eco-routing models for different types/classes of vehicles can be developed.

VED can also be utilized to bridge the gap between the simulation and the reality. The performance of an eco-routing algorithm is heavily dependent on energy consumption models, which are influenced by diverse driving conditions including temperature, time of day, vehicle weights, types of vehicles, power sources, and driving environments; these are available in the dataset. Time-variant speed profiles are also captured, and can be used to increase the accuracy of energy consumption models as discussed in \cite{ref:Eco-routing_Spd}, \cite{ref:Eco-routing_Spd2}. Constructing accurate energy consumption models at intersections is another challenge of current eco-routing algorithms \cite{ref:Eco-routing_Spd}. Utilizing the dataset, which captures abundant real-world driving records at intersections, we may be able to resolve the challenges.

\subsection{Eco-approach and Departure}
An eco-driving concept at signalized intersections, also called as Eco-approach and Departure (EAD), has been studied as summarized in \cite{ref:Review_CAV_Vahidi}. Recent work which studied 609 real-world driving records in free-flow scenarios identified that the fuel saving potential of the EAD technique was on average 40-50\% compared to an average human driver \cite{ref:EAD_Oh}.

The goal of EAD is to resolve the fuel wastage at signalized intersections. EAD aims to obtain energy efficient trajectories of vehicles driving through signalized intersections by utilizing signal timing and phasing information provided over V2X communication.

\begin{figure}[h] 
    \centering
    \includegraphics[width=0.95\linewidth]{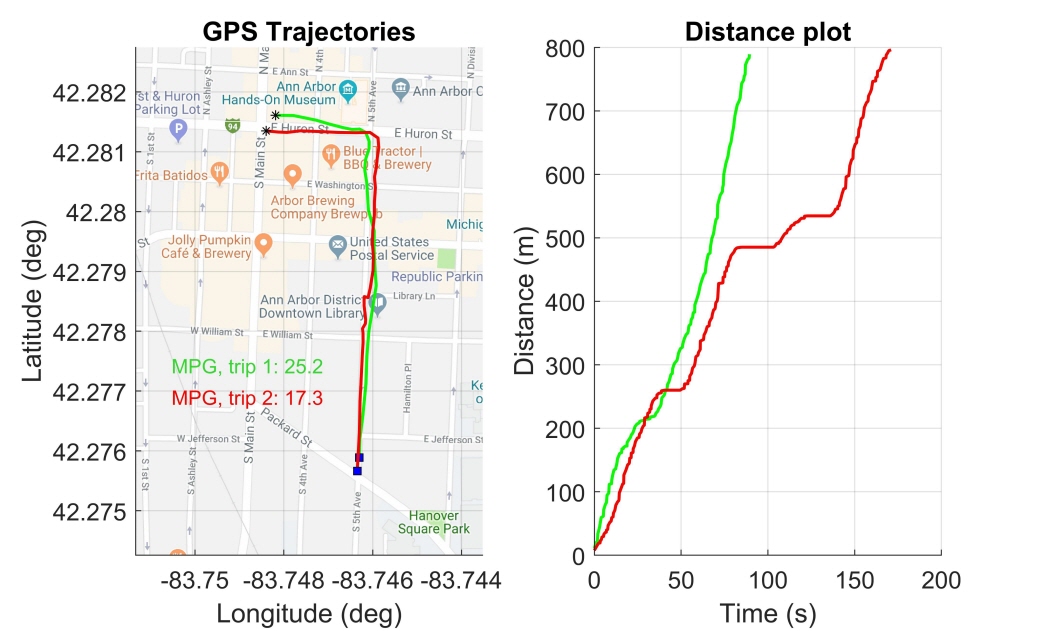}
    \caption{Two distinct trips traveled by a 2013 Honda Accord ICE vehicle. The left plot visualizes the two routes, and the right plot shows the corresponding distance trajectories. MPG for the trip 1 and the trip 2 are 25.2 and 17.3 respectively.}
    \label{Fig:EAD}
\end{figure}

Fig.~\ref{Fig:EAD} demonstrates how the two trips given the same route can have significantly different fuel consumption and travel time. The result indicates that the trip 1 consumed 32\% less fuel than trip 2. It is inferred from the location of signalized intersections and the distance trajectories that trip 2 made several stops at the intersections but trip 1 did not. While surrounding traffic is unobservable and they may have affected the trips, we nevertheless conclude that there exist opportunities to save fuel and travel time by catching the right phase of traffic signals and minimizing stops. \\

\section{Conclusion}

The contributions of the paper are twofold: (1) Introduction of VED, a novel large-scale vehicle energy dataset, and (2) Identification of energy saving opportunities for modern vehicles.

VED contains second-by-second trajectories and time-series energy data of 383 privately owned vehicles consisting of various classes of ICE vehicles, HEVs, PHEVs, and EVs. Driving scenarios captured in VED are diverse, ranging from highways over rural areas to busy downtown areas. The dataset was de-identified to ensure that the personally identifiable information are protected.

To demonstrate how one can utilize the dataset, we conducted a number of case studies and report the following: 
\begin{enumerate}
\item Fuel economy varies significantly depending on the driving environments and travel speeds. MPG of the vehicles in highway were on average 2.92 times higher than in busy inner-city. The most fuel efficient speed, or the highest MPG speed is 18 $(m/s)$.
\item Ambient temperature greatly influences fuel economy of vehicles, and the low temperature has detrimental effect on the fuel economy, resulted in 46\% increased fuel use at $-8^\circ$C compared to $20^\circ$C.
\item Time of the day had impacts on the traffic and fuel economy. Vehicles in rush-hours compared to free-flow hours consumed 67\% more fuel to travel the same distance.
\item HEVs are much more energy efficient than ICE vehicles in urban roads. ICE vehicles consumed on average 287.5\% more fuels to travel the same distance near signalized intersections.
\item Energy use of PHEVs, EVs including battery power and auxiliary power usage can be analyzed from our dataset.
\item The opportunities to save fuel consumption in routing and at signalized intersections were identified and can be achieved by utilizing CAV technologies.
\item Data-driven energy consumption models which consider diverse driving conditions can be built upon the dataset and used for developments of accurate eco-routing and eco-approach and departure algorithms.
\end{enumerate}

In conclusion, VED helps us to better understand behaviors of human drivers and their energy usage, and provides ample research opportunities for CAV technologies and autonomous driving research.

%%%%%%%%%%%%%%%%%%%%%%%%%%%%%%%%%%%%%%%%%%%%
\section*{ACKNOWLEDGMENT}
The authors would like to thank U.S. Department of Energy
for supporting this project. The authors also would like to thank the University of Michigan Transportation Research Institute for data collections, construction and management of the raw database. In particular, our gratitude to Jim Sayer, Mich Rasulis, Scott Bogard, Mary Lynn Buonarosa, and Dillon Funkhouser for fleet recruiting, installation of the data collection devices, database construction and management. \\
%%%%%%%%%%%%%%%%%%%%%%%%%%%%%%%%%%%%%%%%%%%%

\bibliographystyle{IEEEtran}

\end{document}